\def\sec\ond{{\rm s}}

\def\etal{{et al.~}}
\def\ie{{i.e.}}

\documentclass[11pt,emulateapj,apj]{emulateapj}

\shorttitle{Morphology Segregation of Galaxies in the Color-Color Gradient Space}
\shortauthors{Park and Choi}

\begin{document}


\title{Morphology Segregation of Galaxies in the Color-Color Gradient Space}


\author{Changbom Park\altaffilmark{1,2} and Yun-Young Choi\altaffilmark{1,3}}
\altaffiltext{1}{Korea Institute for Advanced Study, Dongdaemun-gu, Seoul 130-722}
\altaffiltext{2}{cbp@kias.re.kr}
\altaffiltext{3}{yychoi@kias.re.kr}


\begin{abstract}
We have found the $u -r$ color versus $g -i$ color gradient space 
can be used for highly successful morphology classification of galaxies 
in the Sloan Digital Sky Survey.  
In this space galaxies form early and late type branches well-separated 
from each other. The location of galaxies along the branches reflects 
the degree and locality of star formation activity, 
and monotonically corresponds to the sequence of morphological subclasses. 
When the concentration index is used together, 
the completeness and reliability of classification reaches about 91\% 
for a training set of SDSS galaxies brighter than $r_{\rm pet}\approx 15.9$. 
At faintest magnitudes ($r_{\rm pet}\approx 17.5$) of the SDSS spectroscopic 
sample, the performance still remains at about 88\%. 
The new classification scheme will help us find accurate 
relations of galaxy morphology with spatial and temporal environments, 
and help one understand the origin of morphology of galaxies.
\end{abstract}



\section{INTRODUCTION}
Unlike stars, galaxies show diverse shapes. 
The belief that the difference in appearance reflects the generic character 
has motivated morphological classification of galaxies. 
Since the 1920s the simple scheme suggested by Hubble (Hubble 1926) based on 
single band images of bright galaxies has been widely adopted. 
The essence of Hubble's scheme and its elaborations (De Vaucouleurs \etal 1991;
Kormendy 1979; De Vaucouleurs 1959; Sandage 1961) is to divide galaxies 
into early and late types. The early types are further divided into 
ellipticals and lenticulars, and the late types into spirals 
(unbarred and barred) and irregulars. Any one who looks directly 
into galaxy images immediately realize that shapes of galaxies are much more 
diverse than this, but also that the simple Hubble sequence catches 
the major features of galaxy morphology. 
As unprecedentedly large sets of digital images of galaxies such as 
the Sloan Digital Sky Survey (SDSS, full details of the SDSS are 
available at http://www.sdss.org. York \etal 2000) data are available, 
time is ripe for understanding the origin of morphology of galaxies. 
In order to find the accurate relationship of galaxy morphology 
with other physical properties it is required to define morphological 
types more objectively and to use morphology subsets as homogeneous as possible.
The endeavors to develop an objective, automatic and highly successful morphology 
classifier applicable to large libraries of digital images 
have resulted in various schemes (Yamauchi \etal 2005; Blanton \etal 2003$a$; 
Abraham, van den Bergh, \& Nair 2003; Strateva \etal 2001; Shimasaku \etal 2001; 
Abraham \etal 1996; Doi, Fukugita, \& Okamura 1993).
This work further advances this topic. 

The philosophy of our morphological classification is to develop a 
scheme applicable down to the faint magnitude limit ($r_{\rm pet} \approx 17.77$ where 
$r_{\rm pet}$ is the Petrosian magnitude after Galactic extinction correction) 
of the spectroscopic sample of the SDSS using only photometric information. 
The fundamental features of galaxy morphology still useful 
for galaxies near the faint limit are the surface brightness 
and color profiles.
Their mean levels are measured by mean surface brightness and integrated color.
Their radial variation scaled to a characteristic size
can be measured by the Sersic or concentration index, 
and color gradient, for example. 
Therefore, in principle, the morphology classification of galaxies should be made
at least in 4-dimensional parameter space. 
The surface brightness at long wavelength bands represents the stellar 
mass distribution, and the color tells about the recent star formation history.
A difficulty in dividing galaxies into early and late types
using the surface brightness profile alone lies in the fact that 
most spirals consist of both bulge and disk. 
There is an unavoidable confusion between the bulge-only and 
the bulge-plus-disk systems when only the surface brightness information 
in a single band is used for classification. 
Contamination in the early and late type subsets separated 
by using the concentration index, 
for example, is typically about 20\% (Yamauchi \etal 2005; Shimasaku \etal 2001). 
The surface texture is of limited usage because it is sensitive to seeing 
and is simply lost for small faint galaxies 
($r_{\rm pet} > 16.0$ in the case of SDSS; Yamauchi \etal 2005).  
On the other hand, the information in color reveals the additional 
generic difference between early and late type galaxies 
which have experienced different star formation histories. 
Strateva \etal (2001) has found that the integrated 
(observer frame) $u^{*}-r^{*}$ color (asterisks are attached to the 
SDSS photometry based on the photometric equations used through 
the Early Data Release) of the SDSS Main galaxies (Stauss \etal 2002) 
shows a bimodal distribution. 
However, they have shown that, when divided at $u^{*}-r^{*} = 2.22$, the 
early (E/S0/Sa) and late type (Sb/Sc/Irr) subsets have significant 
contaminations reaching about 30\% for a sample with visually 
identified morphological types. 
In this paper we extend their work by incorporating color gradient
and the concentration index as additional dimensions of the classification
parameter space.

\section{THE TRAINING SAMPLE}

We use a training set of 1982 galaxies to quantify the performance of our morphology
classification scheme. 
The training set consists of two samples. 
The first contains $981$ SDSS galaxies with $14.5 < r_{\rm pet} \leq 15.0$ 
after Galactic extinction correction whose morphological types are 
assigned by the authors based on the color images retrieved by the 
SDSS Image List Tool (http://cas.sdss.org/astro/en/tools/chart/list.asp). 
Galaxies are divided into six types: E/S0 (smooth brightness and color profile),
 S0 (disk with features, but no spiral arm), 
AE (E or S0 with abnormal brightness or color distribution), 
S (spiral arm, disk with dust lane), AS (spirals with large distortion in shape),
I (irregulars). The first three correspond to the early type, and 
the rest is the late types. The second training set is Fukugita \etal's 
catalog (in preparation; see also Nakamura \etal 2003) 
listing $1,875$ galaxies with $r_{\rm pet}^*\leq15.9$ 
(after Galactic extinction correction) and with visually identified 
morphological types from $T=0$ to $6$ at half integer steps.
 Among them we use a subset of $1183$ galaxies that are 
listed both in the Main Galaxy and the spectroscopic sample and 
are not too much contaminated by foreground stars. 
By comparing the morphological types of $182$ galaxies common 
in our sample and Fukugita \etal's catalog, 
we have found that the early and late types divide at $T = 1.5$ ($T = 1$
and 2 correspond to Hubble Type S0 and Sa, respectively) 
in the latter catalog.

\section{MORPHOLOGY CLASSIFICATION SCHEME}

We use the color versus color gradient space as the major morphology classification tool, 
and use the concentration index as an auxiliary parameter.
In the color-color gradient space spirals tend to 
separate from the region clustered by ellipticals and lenticulars 
as majority of spirals exhibit significant color gradient.
As a measure of color gradient, 
we adopt the difference in $^{0.1}(g -i)$ 
color of the region with $R < 0.5 R_{\rm pet}$  from that of
the annulus with $0.5 R_{\rm pet} < R < R_{\rm pet}$, where $R_{\rm pet}$ 
is the Petrosian radius and $^{0.1}(g -i)$ is a rest frame $g-i$ color 
$K$-corrected 
to the redshift of 0.1 (hereafter, $g -i$ means $^{0.1}(g -i)$),
and $K$-correction of magnitude and color is calculated based on the study 
of Blanton et al. (2003$b$). Negative color difference means bluer outside.
Model magnitude is used for color.  It is the magnitude calculated from the best-fit 
model profile obtained by fitting the de Vaucouleur and exponential profiles to the galaxy image. 
We have chosen the $g$- and $i$-bands to estimate color gradient 
because they are widely separated in wavelength across the 4,000-$\AA$ break, 
and have $S/N$ ratios higher than the $u$- and $z$-bands. 
Choice of other bands results in nosier color gradient estimates 
at faint magnitudes, or less successful separation of galaxies into early 
and late types. 
Other measures of color gradient like the linear slope of the 
radial $g -i$ color profile and the $g -i$ color difference 
between the region with $R < R_{50}$ and the annulus with $R_{50} < R < R_{90}$,
are also calculated for comparisons. 
$R_{50}$ and $R_{90}$ are the radii from
the center of a galaxy containing 50 and 90\% of the Pertrosian flux.
We have found that the clump of the normal early type galaxies 
in the color-color gradient space is less tight for the latter measures
at fainter magnitudes $r_{\rm pet}>16$.

Throughout this paper we use AB magnitudes converted from SDSS magnitudes. 
The $g$- and $i$-band atlas images and basic photometric parameters 
of individual galaxies are retrieved from the SDSS Data Release 3 (DR3) 
at Princeton (http://photo.astro.princeton.edu). 
We have generated a big set of Sersic model images convolved
with the PSF of various sizes. These images are used to find 
a best-fit Sersic model with true Sersic index and true axis ratio for an
observed image of a given size of the PSF. The fitting is made at radii larger
than $0.2 R_{\rm pet}$ to avoid the central region whose profile is much affected by seeing.
Then the best-fit Sersic models in $g$- and $i$-bands are convolved
with the PSF of the same size, and the effects of seeing
on the concentration index and color difference are estimated.
We have confirmed the seeing-corrected concentration index and
color difference show no apparent dependence on seeing.
We use the elliptical annuli in all our parameter calculations 
to take into account flattening or inclination of galaxies. 
The position angle and axis ratio used to define the elliptical annuli 
and to calculate the radial surface brightness profile are isophotal 
ones measured from the $i$-band image. 
The Petrosian radii in our analysis are usually larger than those 
in the DR3 catalog which adopted circular annuli.

The photometric parameters we use for morphology classification are 
the $u -r$ color, color difference $\Delta (g -i)$, 
and the (inverse) concentration index $c$ in the $i$-band,
where $c \equiv R_{50} / R_{90}$.
Here, the concentration index is used as a complementary parameter to discriminate 
red disky spirals from early type galaxies.
The $i$-band image is used to measure the index because the
image at longer wavelengths is expected to better represent the stellar mass
distribution.  All parameters are seeing-corrected as described above.

\begin{figure*}
\plotone{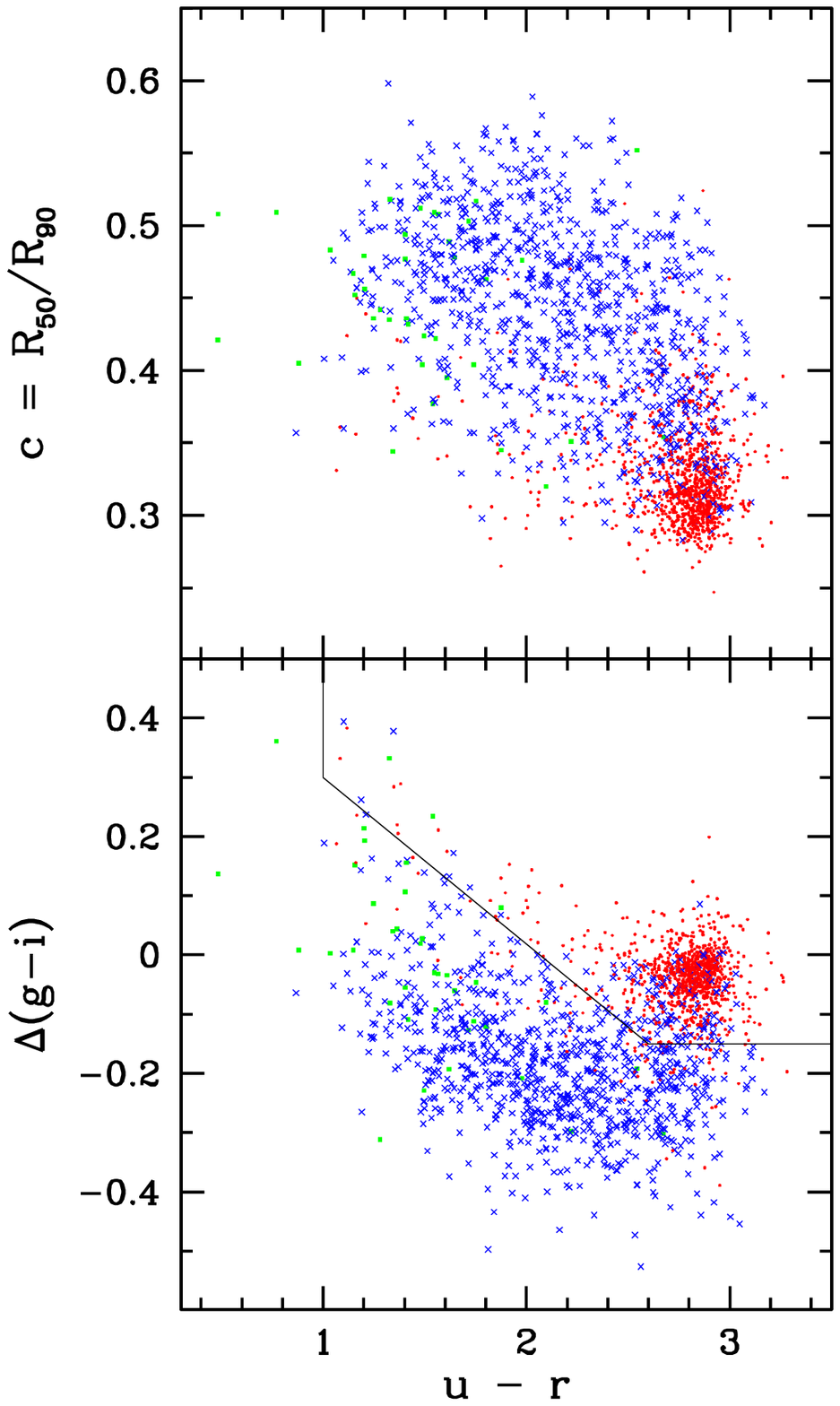}
\caption{Distributions of the $1982$ galaxies in the morphology 
training set in the $u-r$ color versus the (inverse) concentration 
index space (upper panel) and in the $u-r$ color versus the $\Delta (g-i)$ 
color difference space (lower panel). 
Red dots are early types (E/S0, S0, AE), blue crosses are spirals (S, AS), 
and green squares are irregulars (I).} \label{fig1}
\end{figure*}

Figure~\ref{fig1} shows the distribution of galaxies 
of this training set in the $u -r$ versus $\Delta (g -i)$ space and 
in the $u -r$ versus concentration index space. 
The early (E/S0) and late types (Sa to Sd) determined 
by visual inspection are marked as circles (red) and crosses (blue), respectively. 
Squares (green) are irregulars. In our automated scheme classified as early 
type galaxies are those lying above the boundary lines passing through 
the points $(3.5, -0.15)$, $(2.6, -0.15)$ and $(1.0, 0.3)$ 
in the $u -r$ versus $\Delta (g -i)$ space. They are also required to 
have $c < 0.43$. The rest are classified as late types. 
The completeness of this classification scheme reaches $91.3\%$ 
for early types and $90.1\%$ for late types. 
The reliabilities are $90.1\%$ and $91.4\%$, respectively. 
The completeness, $C$, is the fraction of galaxies of a given type 
that are successfully selected from the original sample 
by the classification scheme. 
The reliability, $R$, is the fraction of galaxies of the desired type 
from the selected subsample.
The $9 \sim 10\%$ failure is due to a few red passive spirals, galaxies 
with companions or foreground stars, and to incorrect visual classification 
(for edge-on objects in particular). The parameter $c$ does only an auxiliary 
role improving the results by a few percents. 

We estimate the completeness and reliability of our morphology classification
is about $88\%$ near $r_{\rm pet} = 17.5$, which is close to the faint end 
of the spectroscopic sample of the SDSS.
The performance of our classification scheme slowly degrades at $r_{\rm pet}\geq16$
because it becomes harder to measure the color difference and 
concentration index as the size of galaxies relative to the CCD pixel 
and the FWHM of the PSF decreases. 
To estimate the performance of our classification scheme 
at fainter magnitudes we redshift the galaxies in the training sample 
($r_{\rm pet} = 14.5 \sim 15.0$) with morphological types assigned by us. 
We calculate the new redshift at which a galaxy appear dimmer 
by a desired magnitude adopting a cosmology with density parameters
$\Omega_{\rm m} = 0.27$ and $\Omega_{\rm \Lambda} = 0.73$. 
The reduction factor in angular size is calculated 
and images are binned down (Giavalisco et al. 1996). 
The resulting images are convolved with the PSF 
in such a way that they have the seeing effects equal to those of the original ones. 
Noises are not added anew. Following this procedure we generated three 
mock morphology samples of $981$ galaxies in half magnitude bins from 
$r_{\rm pet} = 16.0$ to $17.5$. The distributions of these redshifted galaxies 
in the color-color gradient space and in the color-concentration index 
space are very similar with those of three samples of randomly-drawn $1,000$ SDSS 
galaxies with actual magnitudes in the same bins, thus justifying our procedure.
We determine the classification criteria at fainter magnitudes 
using these redshifted morphology samples. 
Results are summarized in Table~\ref{table1}.
Table 1 says that our classification
boundaries in the color-color gradient space hardly change even at
magnitudes down to $r_{\rm pet} = 17.5$. The only major change is the boundary 
at the reddest colors (the line connecting to the point P3). At fainter magnitudes the clump of the 
normal E/S0's expands vertically in the color-color gradient space 
because the measured color-differences have more errors. The classification 
boundary at the reddest colors has been chosen to include more ellipticals
taking into account this fact. 
We think the $K$-correction of color in this reddest color range 
has been accurately estimated. Our $K$-correction might have more errors 
for intermediate-type spirals for which the spectra of the central part of galaxies
are less representative of the total color. But in this case our classification is hardly affected 
because these galaxies are blue, have large color gradients, and are located far from 
the classification boundaries. 

\section{DISCUSSION}

Inspection of location of galaxies in the color-color gradient space leads 
us to classification beyond the dichotomous division into early and late types.
Normal ellipticals and lenticulars are very homogeneous systems
whose light is dominated by old Population II stars. 
They strongly concentrate within a spot centered at $(2.82, -0.04)$ in the $u -r$ and $\Delta (g -i)$ plane. Dispersion is mainly due to companions 
or foreground stars. They show very weak but definite color gradient 
(\ie, outside is bluer). We have discovered a trail of early types 
containing about 10\% of early type galaxies toward the left hand direction 
$(u -r  < 2.5)$ from the concentration of normal early types and then upwards. 
The galaxies in this trail are bluer than normal ones, often 
show emission lines, and occasionally show star burst activity near the center.
They tend to be more centrally concentrated than spirals at the same color.
We have found that most $E+A$ galaxies 
(Dressler \& Gunn 1983) listed in Yamauchi \& Goto (2005) are located in this trail.

\begin{figure*}
\epsscale{0.75}
\plotone{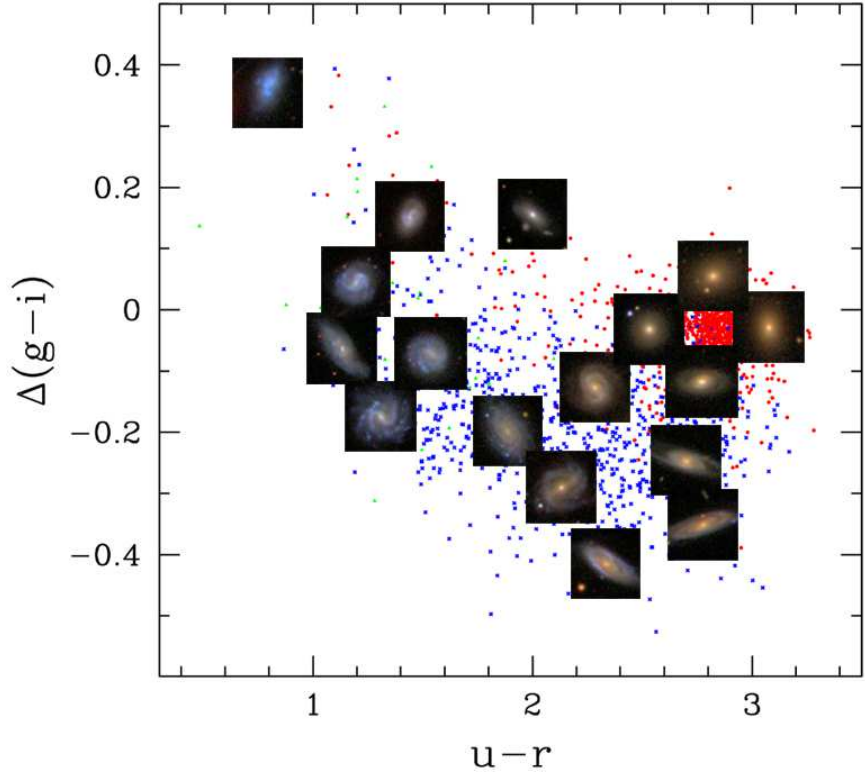}
\caption{A picture showing the continuous changes in the appearance of galaxies 
along the spiral and elliptical branches. 
Note that the star forming region extends from the outskirt of disk 
toward the center as one moves from red to blue spirals.} \label{fig2}
\end{figure*}

\begin{figure*}
\epsscale{0.75}
\plotone{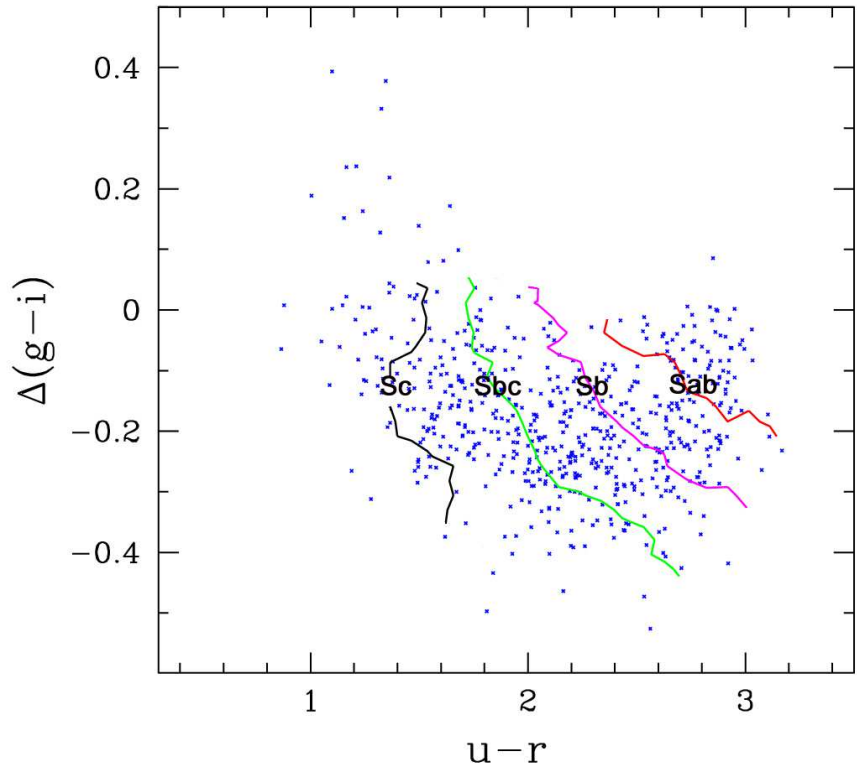}
\caption{Distribution in the $u-r$ color versus the $\Delta (g-i)$ plane 
of $582$ spiral galaxies in the training set taken from 
Fukugita et al.'s morphology sample which lists subclasses for spirals. 
Superposed are the iso-subclass contours calculated 
by averaging the listed subclasses within the ellipse of axis 
lengths of $0.6$ and $0.2$ in the $u-r$ and the $\Delta (g-i)$ directions, 
respectively. Contour levels are $T = 2.5$ (Sab), $3$ (Sb), $3.5$ (Sbc), 
and $4$ (Sc) from right to left.}\label{fig3}
\end{figure*}

On the other hand, a spiral branch is extended downward of the cluster of 
normal early types. Sa type spirals often start to have star forming zones 
at the outskirt of disk. Their integrated color is still dominated by the 
red light from their bulges, but they begin to show color gradient. 
As the star formation activity becomes stronger, 
color gradient increases first and then starts to decrease as the star
forming region expands toward the center. Very late type spirals and most irregulars 
at the upper left corner are very blue and have inverted color 
gradient (\ie, center is bluer) since they usually show strong star 
burst near the center. The spiral branch meets the trail of early types there. 
The major advantage of morphology classification in the color-color gradients 
space is that the blue early type galaxies are separated 
from the spiral galaxies because the former tend to have nearly constant 
color profile or bluer cores while the latter tend to have red bulge plus 
blue disk structure. The spiral branch is a sequence of locality of star 
formation activity and cold gas and dust concentration as illustrated 
in Fig.~\ref{fig2}. The position of a spiral along the branch 
has a strong correlation with its morphological subclass as shown 
in Fig.~\ref{fig3}. 
After making early-late type division, 
one can further divide galaxies into subclasses based on their location 
along the early and late type branches. 
For example, late type galaxies can be divided into $L1$, $L2$, and $L3$ 
subclasses which group galaxies lying above the line $\Delta (g -i) = 0.8 - 0.4 (u -r)$, between this line and the line $\Delta (g -i) = 1.6 - (u -r)$, 
and left of the latter line, respectively. 
These subclasses roughly correspond to Sa-Sb, Sb-Sc, and Sc-Sd/Irr, respectively. 
Dispersion of the spiral branch is partly due to internal extinction 
which tends to make $u -r$ and $\mid\Delta (g -i)\mid$ increase.
Dependences of $u -r$ and $\mid\Delta (g -i)\mid$ on axis ratio of galaxies
indicate that the effects of internal extinction are strongest 
for intermediate type spirals. 

Our new morphology classification scheme takes into account history 
and locality of star formation as well as stellar mass distribution. 
It is easy to implement the scheme for multi-band surveys with 
wide spectrum coverage. 
We expect that the strong concentration of early types and 
the general shape of the spiral branch in the color versus 
color gradient space are still maintained for galaxies at high redshifts. 
Any change in their location and shape will be colorful evidence 
for evolution of star formation activity and cold gas infall into galaxies 
with different morphology.

\acknowledgments
CBP thanks Michael Vogeley for collaboration on parts of this work and
for invitation to Drexel University when this work was started.
We also thank Yasushi Suto for suggesting us to use Fukugita et al.'s
catalog.
We thank the anonymous referee for helpful comments.
This work is supported by the Korea Science and Engineering Foundation 
(KOSEF) through the Astrophysical Research Center for the Structure 
and Evolution of Cosmos (ARCSEC).

\clearpage


\begin{table}[b]\footnotesize
\begin{center}
\caption{Classification criteria}\label{table1}
\begin{tabular}{ccccc}
\hline\hline
Magnitude bins 	&$14.5\sim 16.0$&	$16.0\sim 16.5$&	$16.5\sim 17.0$&	$17.0\sim 17.5$\\
\hline
Classification Criterion Parameters\\
$P1$	&$(1.00, 0.30)$&$(1.00, 0.30)$&$(1.00, 0.30)$&$(1.00, 0.30)$\\
$P2$	&$(2.60, -0.15)$&$(2.65,-0.18)$&$(2.65,-0.18)$&$(2.70,-0.18)$\\
$P3$	&$(3.50, -0.15)$&$(3.50, -0.15)$&$(3.50, -0.25)$&$(3.50, -0.35)$\\
$c_{\rm cut}$&0.43&	0.45 & 0.47&0.48\\
\hline
Completeness (C) and Reliability (R)\\
C(Elliptical)	&0.913&	0.883&	0.872&	0.883\\
R(Elliptical)	&0.901&	0.892&	0.882&	0.881\\
C(Spiral)	&0.901&	0.902&	0.892&	0.890\\
R(Spiral)	&0.914&	0.893&	0.883&	0.892\\
\hline
\end{tabular}
\end{center}
{
The three points, $P1$, $P2$, and $P3$, define the lines dividing 
the SDSS galaxies into the early and late types in the $u-r$ versus 
$\Delta (g-i)$ plane. The early types are also required to have $c < c_{\rm cut}$.}
\end{table}

\end{document}